\documentclass{jpp}
\usepackage{graphicx}
\usepackage[utf8]{inputenc}
\usepackage[T1]{fontenc}
\usepackage{amsmath}
\usepackage{xcolor}
\usepackage[normalem]{ulem}

\shorttitle{Effect of stratification on resistive instabilities}
\shortauthor{Faisal Sayed, Anna Tenerani and Richard Fitzpatrick}

\title{Resistive instabilities of current sheets in stratified plasmas with a gravitational field}

\author{Faisal Sayed\aff{1,2}
  \corresp{\email{faisal\_sayed@utexas.edu}},
  Anna Tenerani \aff{1}
 \and Richard Fitzpatrick\aff{1}}

\affiliation{
\aff{1} Institute for Fusion Studies, Department of Physics, The University of Texas at Austin, Austin, TX 78712, USA

\aff{2} Department of Physics, Faculty of Science, Assiut University, Asyut 71516, Egypt
}
\begin{document}

\maketitle
\begin{abstract}
Magnetic reconnection can develop spontaneously via the tearing instability, often invoked to explain disruptive instabilities in fusion devices, solar flares, the generation of periodic density disturbances at the tip of helmet streamers, and flux transfer events at the Earth’s dayside magnetopause. However, in many such environments the presence of gravity, magnetic field curvature or other forms of acceleration often result in situations of a heavy-over-light plasma in an effective gravitational field with an embedded current sheet. This paper studies the linear stability of a slab current sheet with respect to reconnecting modes in the presence of a density gradient under the effect of a constant gravitational acceleration. We show that the presence of stratification and gravity modify the properties of the tearing mode instability both in the case of favorable and unfavorable stratification. Favorable stratification suppresses reconnection while unfavorable stratification strongly destabilizes the tearing mode. Furthermore, we show that the classical constant-$\psi$  regime effectively does not exist, even for weak unfavorable stratification, for $S \gg 1$. Instead, the gravity-modified tearing progressively transitions into the G‑mode, which is a gravity-driven reconnecting mode with a growth rate scaling as $S^{-1/3}$. As a consequence, unfavorable stratification only permits rapidly reconnecting modes.   
\end{abstract}

\section{Introduction}
\label{sec:intro}
Magnetic reconnection is thought to be the dynamical mechanism responsible for many explosive phenomena observed  in laboratory, space, and astrophysical plasmas \citep{1,pucci2020onset,ji2022preface}. Through magnetic topology reconfiguration, reconnection converts stored magnetic energy into plasma heating, particle acceleration, and bulk flows. This mechanism is central to energy release events in magnetic confinement fusion devices (e.g., tokamak sawtooth crashes \citep{17}), solar flares and coronal mass ejections in the solar corona \citep{18}, geomagnetic storms and substorms within planetary magnetospheres \citep{burch2016magnetic,louarn2015magnetic}, and high-energy phenomena in astrophysical environments such as accretion disks and relativistic jets \citep{romanova1992magnetic,khiali2016high}.

The tearing mode instability, first analyzed by Furth, Killeen, and Rosenbluth (1963) \citep{6}, is a fundamental mechanism through which magnetic reconnection can develop within current sheets in the resistive magnetohydrodynamic (MHD) framework. Over the past several decades, the tearing mode has been investigated extensively under increasingly complex plasma conditions.  Studies have explored its properties  with equilibrium shear flows and/or viscosity \citep{9,42,43,44,45,dahlburg1997evolution,tenerani2015tearing}, incorporated Hall and two-fluid effects \citep{10,pucci2017fast,shi2020oblique}, and extended to partially ionized plasmas \citep{paris1973tearing,12,pucci2020tearing} and to collisionless or kinetic regimes \citep{porcelli1991collisionless,daughton2005kinetic,del2016ideal}.  

However, in many natural and laboratory environments reconnection can occur in stratified plasmas in the presence of gravity, magnetic curvature, or other effective accelerations, sometimes leading to heavy-over-light plasma configurations embedded within current sheets. At Earth’s magnetopause, for example, rapid variations in solar wind dynamic pressure accelerate the boundary across the sharp density gradient separating the dense magnetosheath from the tenuous magnetosphere, producing an effective gravitational force that can drive Rayleigh-Taylor modes \citep{15,16} and potentially affect reconnection during southward interplanetary magnetic field conditions. Similar conditions arise at the heliopause, where both acceleration of the heliospheric interface toward the interstellar medium and charge-exchange interactions between interstellar neutrals and heliosheath plasma can generate an effective gravity antiparallel to the density gradient, making the interface unstable to Rayleigh-Taylor modes \citep{31,32,33,34,ruderman2024heliopause}. In the laboratory, magnetic confinement devices such as tokamaks and stellarators are also susceptible to interchange-type instabilities \citep{kruskal1954some} and other gravity-driven resistive instabilities  that can develop in current sheets \citep{6, 23, johnson1963effect, 39, 40}. These early works however relied on restrictive assumptions --- such as constant-$\psi$, weak gravity, weak magnetic shear, or asymptotic wavelength limits --- thereby limiting their ability to capture how tearing modes are modified by gravity and how they transition to gravity-driven instabilities.

In this study, we investigate the effects of favorable and unfavorable stratification on the linear stage of reconnecting instabilities in a plane slab current sheet by including a gravitational acceleration acting normal to the current layer. Here, favorable and unfavorable stratification refer to cases in which the density gradient is parallel or antiparallel, respectively, to the gravitational acceleration. Although configurations with a negative density gradient along the gravitational field direction (unfavorable stratification) can be ideally unstable to interchange or Rayleigh-Taylor modes, we restrict our attention to regimes in which stratification is sufficiently weak to satisfy the Suydam stability criterion \citep{24,gerwin1979some}, a necessary condition for ideal stability in a sheared pinch configuration. 

With uniform resistivity, it has been shown that a plane sheet pinch is unstable to two resistive instabilities under the constant-$\psi$ approximation \citep{6}: a long-wavelength tearing mode, dominant when gravity is negligible, with a growth rate scaling with the Lundquist number as $S^{-3/5}$, and a short-wavelength gravity-driven mode (the G-mode) with a growth rate scaling  as $S^{-1/3}$. This gravity-driven resistive mode was further extended to stellarator-like geometry in weakly sheared magnetic configurations \citep{johnson1963effect,23,39} and generalized to non-localized modes manifesting as twisted convective rolls \citep{39}. Recently, Hopper et al. \citep{3} investigated the effect of favorable stratification on the linear growth rate of the tearing mode in slab geometry using both analytical and numerical approaches. They found that increasing the stabilizing density gradient suppresses the growth of tearing perturbations. Here, we follow their approach in solving the resistive layer equations assuming a linear density profile, but extend their analysis to include unfavorable stratification, allowing for  gravity-driven instabilities. We derive the general dispersion relation that describes the continuous transition from the tearing-dominated regimes to G-modes and examine how the growth rate depends on both the stratification strength (favorable and unfavorable) and the Lundquist number. In addition, we numerically investigate a more realistic equilibrium density profile with a smooth transition between two plasma regions to assess the sensitivity of the growth rate to the density profile.

The plan of the paper is as follows. We will begin in Section \S\ref{sec2} by introducing the set of equations describing the tearing mode and derive layer equations in Sections \S\ref{sec3} and \S\ref{sec4}. In section \S\ref{sec5}, we show the dispersion relation obtained by applying the asymptotic matching technique. In Section \S\ref{sec6} we compare analytical and numerical results and investigate the effects of stratification on the growth rate of the tearing mode. Conclusions of this study are given in Section \S\ref{sec7}. 

\section{Model equations}
\label{sec2}
We start from the resistive MHD equation set that neglects plasma compressibility and incorporates uniform plasma resistivity
\begin{equation}
\label{1}
\frac{\partial \rho}{\partial t} +\nabla\boldsymbol{\cdot}(\rho\boldsymbol{V}) = 0 , 
\end{equation}
\begin{equation}
\label{2}
 \rho\frac{\partial \boldsymbol{V}}{\partial t} + \rho\boldsymbol{V}.\nabla\boldsymbol{V} = - \nabla P + \boldsymbol{J}\times \boldsymbol{B} + \rho \boldsymbol{g} ,
\end{equation}
\begin{equation}
\label{3}
\frac{\partial\boldsymbol{B}}{\partial t} =\nabla \times (\boldsymbol{V}\times\boldsymbol{B})+ \frac{\eta}{\mu_{o}} \nabla^{2}\boldsymbol{B},
\end{equation}
\begin{equation}
\label{4}
\nabla \boldsymbol{\cdot} \boldsymbol{V} = 0,
\end{equation}
where $\rho$, $\boldsymbol{V}$, $P$, $\boldsymbol{B}$  are the density, velocity, scalar pressure, and magnetic field respectively. $\boldsymbol{J} = \nabla \times \boldsymbol{B}/\mu_{0}$ is the current density and $\boldsymbol{g}$ is the gravitational acceleration acting in the negative $x$-direction, and $\eta$ is the resistivity. 

\begin{figure}
    \centering
    \includegraphics[width=0.8\textwidth]{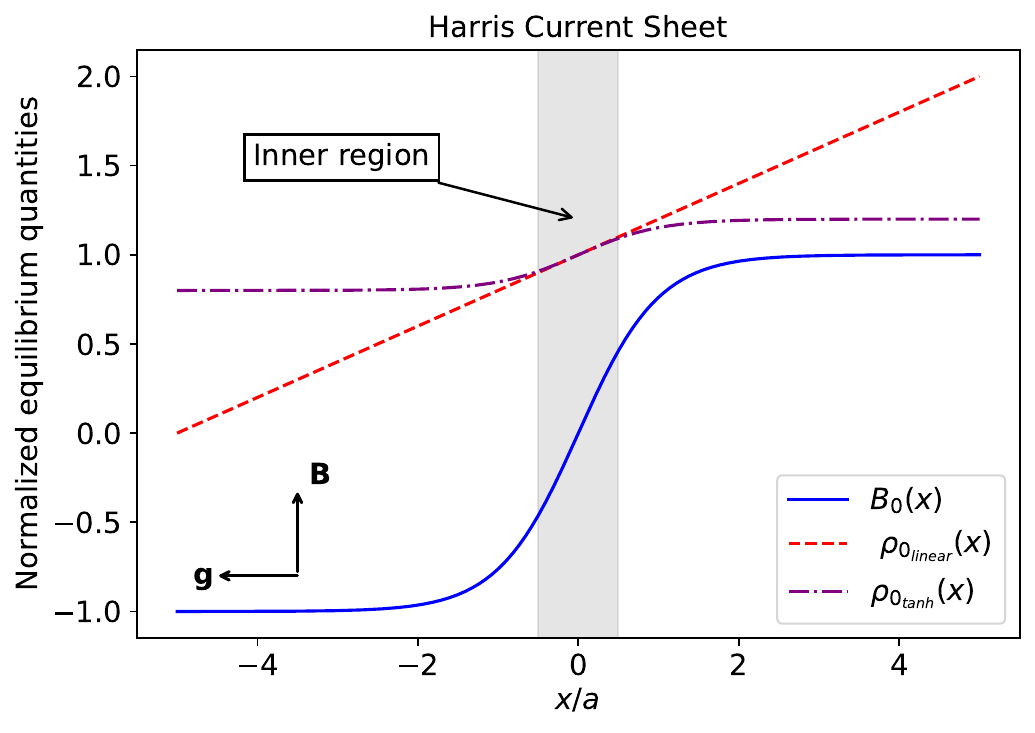}
    \caption{Schematic of the equilibrium configuration within the Harris-type current sheet in the presence of an effective gravitational field, highlighting the equilibrium magnetic field and the two density profiles considered in this work: the linear profile (equation \ref{6}) and the hyperbolic tangent profile (equation \ref{58}).}
    \label{equilibrium}
\end{figure}

Our equilibrium, illustrated schematically in figure \ref{equilibrium}, is characterized by an inhomogeneous plasma embedded in a Harris-type current sheet  \citep{27}, defined by the following in-plane magnetic field and plasma density,
\begin{equation}
\label{5}
\boldsymbol{B}_{0} = B_{0} \tanh \left(\frac{x}{a}\right) \boldsymbol{\hat{y}},
\end{equation}
\begin{equation}
\label{6}
    \rho_{0}(x) = \bar{\rho}_{0}\left(1+\epsilon \frac{x}{a}\right),
\end{equation}
where $a$ is the current sheet half-thickness and $\bar{\rho}_{0}$ is a constant. We assume a weak density inhomogeneity ($\epsilon\ll1$) in order to apply the Boussinesq approximation. Our choice of the linear density profile enables an analytic solution to the outer equation. This profile can be regarded as an approximation near the resonant surface of the more physically realistic tanh density profile introduced later (see equation \ref{58}), since $\tanh(x/a)\approx x/a$ for $|x/a|\ll1$. Consequently, the behavior of the outer solution should be interpreted as a consequence of this modeling choice. The background flow is set to zero everywhere ($\boldsymbol{V}_{0} ={\boldsymbol 0 }$) and the equilibrium is achieved with a background pressure $P_{0}(x)$ such that 
\begin{equation}
\label{7}
    P_0(x) + \frac{B_{0}^{2}}{2\mu_{o}}+\bar{\rho}_{0}\left(x+\epsilon \frac{x^2}{2a}\right)g = P_0(0).
\end{equation}
The configuration described above satisfies the equilibrium conditions of ideal MHD. However, when finite resistivity is included, the magnetic field diffuses on the resistive diffusion timescale $\tau_R=\mu_{o}a^2/\eta$. In this work, we focus on weakly collisional plasmas, where resistivity is assumed to be sufficiently small such that the diffusion rate is negligible compared to the instability growth rate, allowing us to neglect resistive diffusion in the analysis. An exception arises in the favorable stratification asymptotic regime, where we will show that $\gamma\tau_A \sim S^{-1}$ (Section \ref{sec4.3}). In this limit, the instability growth timescale becomes comparable to the resistive diffusion timescale of the equilibrium current sheet, and the assumption of a fixed equilibrium is therefore no longer self consistent. As discussed in Section \ref{sec4.3}, the $S^{-1}$ scaling should thus be regarded as a formal asymptotic result obtained under the assumption of a frozen equilibrium; unless the current sheet is externally maintained, its resistive evolution must be taken into account.

In the context of a stratified atmosphere it is common to adopt entropy, rather than density, as the dynamically advected thermodynamic variable within the Boussinesq approximation. Our work however is intended to describe a different physical regime: an incompressible current sheet embedded between plasma regions of different densities in an effective gravitational field, rather than a self-consistent stratified atmosphere. The motivation is to model environments such as the magnetopause and heliopause, which are characterized by a density gradient between adjacent plasma regions. In such systems, the relevant buoyancy effect arises directly from the density gradient across the interface, analogous to Rayleigh–Taylor/interchange  instability, rather than from an entropy gradient. Accordingly, we adopt an incompressible description in which the equilibrium density profile is prescribed and weakly varying, while compressibility effects are neglected. Density variations are retained only in the gravitational (buoyancy) term and neglected elsewhere to leading order, consistent with Boussinesq ordering. Consequently, the stratification parameter $A$ (introduced later in equation (\ref{parameter_A})) measures the stabilizing or destabilizing effect of an imposed density variation in the presence of gravity, rather than the Brunt–Väisälä response of a stratified atmosphere.

To linearize the system of equations (\ref{1})-(\ref{4}), we first introduce the two scalar stream ($\phi$) and flux ($\psi$) functions such that
\begin{equation}
\label{8}
    \boldsymbol{V} = \nabla\phi\times\boldsymbol{\hat{z}},
\end{equation}
\begin{equation}
\label{9}
    \boldsymbol{B} = \nabla\psi\times \boldsymbol{\hat{z}}.
\end{equation}
We consider small fluctuations of the form $f(x/a)$ $\text{exp} (iky+\gamma t)$, where $\gamma$ and $k$ denote the  growth rate and wave vector of a given mode, respectively. Linearizing the MHD equations around the prescribed equilibrium and introducing these perturbations, we obtain two coupled equations for the normalized flow and magnetic flux perturbations $\hat{\phi}$ and $\hat{\psi}$:
\begin{equation}
\label{10}
\hat{\gamma} S \left(\hat{\psi} -F \hat{\phi}\right) = \left(\hat{\psi}^{''}-\hat{k}^{2} \hat{\psi}\right),
\end{equation}
\begin{equation}
\label{11}
\hat{\gamma}^{2}  \left( \hat{\phi}^{''}- \hat{k}^{2} \hat{\phi}\right) =- \hat{k}^{2} F \left(\hat{\psi}^{''} -\hat{ k}^{2} \hat{\psi} - \frac{F^{''}}{F} \hat{\psi}\right) - \frac{a \mu_{0}}{B_{0}^{2}} g \hat{k}^{2}\hat{\phi} \frac{d\rho_{0}}{d\hat{x}}- \frac{\hat{\gamma}^{2}}{{\bar\rho_{0}}}\hat{\phi}^{'}\frac{d\rho_{0}}{d\hat{x}}.
\end{equation}
In the equations above, a prime denotes differentiation with respect to the normalized variable $\hat{x}=x/a$, $\hat{k}=ka$, and $\hat{\gamma}=\gamma\tau_a$ where $\tau_{a}=a/V_{A}$ is the Alfvén crossing time with $V_{A}=B_0/\sqrt{\mu_0\bar{\rho}_0}$ being the Alfvén speed and $\bar{\rho}_{0}$ is the density at the neutral line ($x=0$). Flow and flux functions are normalized to $\psi = - a B_{0} \hat{\psi}$ and $\phi = i(\frac{\gamma a}{k})\hat{\phi}$. Here, $F=\tanh{(\hat{x})}$ and $S = \mu_{0}aV_{A}/\eta$ is the Lundquist number.

Equation (\ref{11}) has been obtained by taking the $z$-component of the curl of equation (\ref{2}) and then by approximating $\rho_0^{-1}\simeq 1/\bar\rho_0$ to zeroth order in $\epsilon$. We consider the asymptotic limit $\epsilon\rightarrow0$ while holding the stratification parameter, defined in equation \ref{parameter_A}, $A\sim\epsilon \frac{ga}{V^2_A}$ finite and small, $A\ll1$, which implies $\frac{ag}{V^2_A}\sim\epsilon^{-1}$. Under this ordering, the last term in equation (2.11), which is directly proportional to parameter $\epsilon$, vanishes as $\epsilon\rightarrow0$. Whereas, the buoyancy term remains finite through the combination $A=\epsilon\frac{ga}{V^2_A}$. Therefore, the last term can be consistently neglected while a weak but finite buoyancy contribution is retained. Importantly, its omission allows us to solve the eigenvalue problem by imposing specific parity conditions on the eigenfunctions. The resulting equation for the stream function is:
\begin{equation}
    \label{12}
\hat{\gamma}^{2}  \left( \hat{\phi}^{''}- \hat{k}^{2} \hat{\phi}\right) =- \hat{k}^{2} F \left(\hat{\psi}^{''} -\hat{ k}^{2} \hat{\psi} - \frac{F^{''}}{F} \hat{\psi}\right) - A  \hat{k}^{2}\hat{\phi},
\end{equation}
where
\begin{equation}
\label{parameter_A}
A =  \frac{a \mu_{0}}{B_{0}^{2}} g \frac{d\rho_{0}}{d\hat{x}}=\epsilon \frac{a g}{V_{a}^2}  
\end{equation}
represents the stratification parameter and characterizes the strength of stratification, which can be either unfavorable or favorable. According to equation (\ref{parameter_A}), the type of stratification can be controlled through the sign of $\epsilon$ in equation (\ref{6}): $\epsilon < 0$ corresponds to favorable stratification, whereas $\epsilon > 0$ indicates unfavorable stratification. As we mentioned earlier, unfavorable stratification corresponds to the case where the density gradient is antiparallel to the effective gravitational field, resulting in positive values of $A$. On the other hand, favorable stratification refers to the situation where the density gradient and effective gravitational field are aligned, yielding a negative value of $A$. 

The eigenfunctions of equations (\ref{10}) and (\ref{12}) exhibit well-defined parity: the ideal RT mode has the magnetic flux function $\hat{\psi}$ odd and the velocity stream function $\hat{\phi}$ even, whereas the tearing mode has $\hat{\psi}$ even and $\hat{\phi}$ odd. In what follows, we seek solutions to these equations that satisfy the tearing-mode parity. 
\section{Boundary layer approach}
\label{section3}

In this section, we derive the analytic form of the dispersion relation by asymptotically matching the outer and inner region solutions. We begin with the outer layer equations and their solution followed by the solution of the inner layer, and finally we derive the dispersion relation. 

\subsection{Outer region}
\label{sec3}
For $S\gg1$, the dynamics is well described by ideal MHD far from the boundary layer. In such ideal outer region, one can therefore neglect plasma resistivity in equation (\ref{10}) as well as plasma inertia in equation (\ref{12}) and obtain the following ideal equations:
\begin{equation}
    \label{13}
      \hat{\psi}= F\hat{\phi},
\end{equation}
\begin{equation}
    \label{14}
    F\left(\hat{\psi}^{''} -\hat{ k}^{2} \hat{\psi} -\frac{F^{''}}{F} \hat{\psi}\right) + A \hat{\phi} = 0.
\end{equation}
Upon inserting equation (\ref{13}) into (\ref{14}) one can eliminate the flux function and obtain a second order equation for $\hat\phi$: 
\begin{equation}
    \label{15}
    F^{2}\hat{\phi}^{''}+ 2 F F^{'}  \hat{\phi}^{'} +\left(A-\hat{k}^{2} F^{2}\right)\hat{\phi} =0.
\end{equation}
Note that this equation, in the limit $|\hat{x}| \rightarrow \infty$, admits two distinct types of solutions depending on the sign of the coefficient ($A - \hat{k}^{2}$). If $A - \hat{k}^{2} <0$, the solution exhibits an exponentially decaying behavior. In contrast, if $A - \hat{k}^{2}>0$, the solution is instead  oscillatory. Such a non-localized character of the outer solution is an artifact of the linear density profile chosen for analytic tractability, and can be mitigated by imposing a smooth profile of $\rho_{0}(x)$ that asymptotes to constant values for $|\hat x|\rightarrow \infty$. We will consider the latter density profile when discussing our results in Section \ref{sec6}. 

To perform the asymptotic matching between the outer and inner solutions, we first determine the behavior of the outer solution near the resistive layer. To achieve this, we apply the Frobenius expansion technique to obtain a power series solution to equation (\ref{15}) for $\hat |x|\rightarrow0$, where the leading exponent is determined by the corresponding indicial equation. The general solution is then expressed as a linear combination of the leading terms of two linearly independent solutions corresponding to the roots of the indicial equation, and can be written as
\begin{equation}
\label{sol}
\hat{\phi}(\hat{x}) = c_{1}\hat{x}^{-l-\frac{1}{2}} +c_{2} \hat{x}^{l-\frac{1}{2}},
\end{equation}
where $l = \sqrt{\frac{1}{4}-A}$. To ensure a power law behavior near the resistive layer, the condition $A<1/4$ must be satisfied, a requirement that corresponds to Suydam stability condition for our geometry \citep{24}. The constants $c_{1}$ and $c_{2}$ are determined by comparing equation (\ref{sol}) with the complete solution of equation (\ref{15}) in the limit $|\hat{x}|\rightarrow 0$ (see Appendix \ref{AppendixA}). For $\hat{k}^{2}>A$ this leads to
\begin{equation}
    \label{29}
    \hat{\phi}(\hat{x})= \left[a_{0} |\hat{x}|^{-l-\frac{1}{2}}+b_{0} |\hat{x}|^{l-\frac{1}{2}}\right]\,\mathrm{sgn}(\hat{x}),
\end{equation}
where the coefficients are
\begin{equation}
    \label{30}
     a_{0} = \frac{\Gamma(1+\nu)\Gamma(l)}{\Gamma\left(\frac{\nu}{2}+\frac{l}{2}+\frac{5}{4}\right)\Gamma\left(\frac{\nu}{2}+\frac{l}{2}-\frac{1}{4}\right)},
\end{equation}
\begin{equation}
    \label{31}
     b_{0} = \frac{\Gamma(1+\nu)\Gamma(-l)}{\Gamma\left(\frac{\nu}{2}-\frac{l}{2}+\frac{5}{4}\right)\Gamma\left(\frac{\nu}{2}-\frac{l}{2}-\frac{1}{4}\right)}.
\end{equation}
For the case  $A>\hat{k}^{2}$, the solution takes instead the form 
\begin{equation}
    \label{29m}
    \hat{\phi}(\hat{x})=\left[a_{0mod} |\hat{x}|^{-l-\frac{1}{2}}+b_{0mod} |\hat{x}|^{l-\frac{1}{2}}\right]\,\mathrm{sgn}(\hat{x}),
\end{equation}
where the modified coefficients $a_{0mod}$ and $b_{0mod}$ are
\begin{equation}
    \label{32}
     a_{0 mod} = \frac{\Gamma(1+\nu)\Gamma(l)}{\Gamma\left(\frac{\nu}{2}+\frac{l}{2}+\frac{5}{4}\right)\Gamma\left(\frac{\nu}{2}+\frac{l}{2}-\frac{1}{4}\right)}+\frac{\Gamma(1-\nu)\Gamma(l)}{\Gamma\left(-\frac{\nu}{2}+\frac{l}{2}+\frac{5}{4}\right)\Gamma\left(-\frac{\nu}{2}+\frac{l}{2}-\frac{1}{4}\right)},
\end{equation}
\begin{equation}
    \label{33}
     b_{0mod} = \frac{\Gamma(1+\nu)\Gamma(-l)}{\Gamma\left(\frac{\nu}{2}-\frac{l}{2}+\frac{5}{4}\right)\Gamma\left(\frac{\nu}{2}-\frac{l}{2}-\frac{1}{4}\right)}+\frac{\Gamma(1-\nu)\Gamma(-l)}{\Gamma\left(-\frac{\nu}{2}-\frac{l}{2}+\frac{5}{4}\right)\Gamma\left(-\frac{\nu}{2}-\frac{l}{2}-\frac{1}{4}\right)}.
\end{equation}
In these coefficients $\nu=\sqrt{\hat{k}^2-A}$. A detailed derivation of equations (\ref{29}) and (\ref{29m}) is provided in Appendix \ref{AppendixA}.
\subsection{Inner region}
\label{sec4}
In the inner layer, where perturbed magnetic and velocity fields exhibit sharp gradients and where reconnection occurs, resistive effects must be included. Due to these large gradients, we can assume that $\partial^2/\partial\hat{x}^{2} \gg \hat{k}^{2}$. In this way, the inner region equations become
 \begin{equation}
\label{34}
\hat{\gamma} S \left(\hat{\psi} -\hat{x} \hat{\phi}\right) = \hat{\psi}^{''},
\end{equation}
\begin{equation}
    \label{35}
\hat{\gamma}^{2} \hat{\phi}^{''} =- \hat{k}^{2} \hat{x} \hat{\psi}^{''}   - A\hat{k}^{2}\hat{\phi}.
\end{equation}
This set of equations is analytically solvable in Fourier space as the transformation with respect to the variable $\hat x$ reduces the original fourth-order ODE system to a system of second-order differential equations in the variable $\hat{\theta}$. To achieve asymptotic matching with the outer region solution, the inner solution must be determined in the limit $\hat x\rightarrow\infty$ in configuration space, or  $\hat\theta\rightarrow 0$ in Fourier space. In this limit, the Fourier-transformed stream function $\tilde{\phi}(\hat{\theta})$ is 
\begin{equation}
    \label{48}
      \tilde{\phi}(\hat{\theta})=\left[\tilde{a}_{0}|\hat{\theta}|^{l-\frac{1}{2}}+\tilde{b}_{0}|\hat{\theta}|^{-l-\frac{1}{2}}\right]\,\mathrm{sgn}(\hat{\theta}),
\end{equation}
where the coefficients $\tilde{a}_{0}$ and $\tilde{b}_{0}$ are given by the following forms
\begin{equation}
    \label{49}
    \tilde{a}_{0}=\left({\frac{\sqrt{\hat{\gamma}/S}}{\hat{k}}}\right)^{\frac{l}{2}-\frac{1}{4}}\frac{2\omega}{(\omega-l+\frac{1}{2})}\frac{\Gamma(-l)}{\Gamma\left(\frac{(\omega-l)^{2}-\frac{1}{4}}{4\omega}\right)},
\end{equation}
\begin{equation}
    \label{50}
    \tilde{b}_{0}=\left({\frac{\sqrt{\hat{\gamma}/S}}{\hat{k}}}\right)^{-\frac{l}{2}-\frac{1}{4}}\frac{2\omega}{(\omega+l+\frac{1}{2})}\frac{\Gamma(l)}{\Gamma\left(\frac{(l+\omega)^{2}-\frac{1}{4}}{4\omega}\right)},
\end{equation}
with $\omega = \sqrt{\frac{\hat{\gamma}^{3}S}{\hat{k}^2}}$. Appendix \ref{AppendixB} contains the full derivation of equation (\ref{48}). 
\subsection{Dispersion relation}
\label{sec5}
To obtain the dispersion relation, we asymptotically match the outer and inner solutions in configuration space by using the asymptotic expressions provided in the previous sections.  Thus, we first perform an inverse Fourier transform of the inner solution (\ref{48})  into $\hat{x}$ space
\begin{equation}
        \label{53}
     \hat{\phi}(\hat{x})=\left[\bar{a}_{0}|\hat{x}|^{-l-\frac{1}{2}}+\bar{b}_{0}|\hat{x}|^{l-\frac{1}{2}}\right]\,\mathrm{sgn}(\hat{x}),
\end{equation}
where the coefficients $\bar{a}_{0}$ and $\bar{b}_{0}$ are related to the coefficients $\tilde{a}_{0}$ and $\tilde{b}_{0}$, given by equations (\ref{49})-(\ref{50}), as follows,
\begin{equation}
    \label{54}
  \bar{a}_{0} = \frac{2}{i \pi}\left({\frac{\sqrt{\hat{\gamma}/S}}{\hat{k}}}\right)^{\frac{l}{2}-\frac{1}{4}}\frac{\omega \cos\left(\frac{\pi}{2}\left(\frac{1}{2}-l\right)\right) }{\left(\omega-l+\frac{1}{2}\right)}\frac{\Gamma(\frac{1}{2}+l)\Gamma(-l)}{\Gamma\left(\frac{(\omega-l)^{2}-\frac{1}{4}}{4\omega}\right)},
\end{equation}
\begin{equation}
    \label{55}
  \bar{b}_{0} = \frac{2}{i \pi}\left({\frac{\sqrt{\hat{\gamma}/S}}{\hat{k}}}\right)^{-\frac{l}{2}-\frac{1}{4}}\frac{\omega\cos\left(\frac{\pi}{2}\left(\frac{1}{2}+l\right)\right)}{\left(\omega+l+\frac{1}{2}\right)}\frac{\Gamma(\frac{1}{2}-l)\Gamma(l)}{\Gamma\left(\frac{(\omega+l)^{2}-\frac{1}{4}}{4\omega}\right)}.
\end{equation} 
Next,  we match the coefficients of the two leading-order terms in equations (\ref{29}) and (\ref{53}) when $A-\hat{k}^2<0$, yielding the following dispersion relation:
\begin{equation}
\begin{split}
\label{56}
&\left({\frac{\sqrt{\hat{\gamma}/S}}{\hat{k}}}\right)^{l}\frac{(\omega+l+\frac{1}{2})}{(\omega-l+\frac{1}{2})}\frac{\Gamma(-l)}{\Gamma(l)}\frac{\Gamma\left(\frac{(\omega+l)^{2}-\frac{1}{4}}{4\omega}\right)}{\Gamma\left(\frac{(\omega-l)^{2}-\frac{1}{4}}{4\omega}\right)}=\\
&\frac{\cos\left(\frac{\pi}{2}(\frac{1}{2}+l)\right)\Gamma(\frac{1}{2}-l)}{\cos\left(\frac{\pi}{2}(\frac{1}{2}-l)\right)\Gamma(\frac{1}{2}+l)}\frac{\Gamma(l)}{\Gamma(-l)}\frac{\Gamma(\frac{\nu}{2}-\frac{l}{2}+\frac{5}{4})\Gamma(\frac{\nu}{2}-\frac{l}{2}-\frac{1}{4})}{\Gamma(\frac{\nu}{2}+\frac{l}{2}+\frac{5}{4})\Gamma(\frac{\nu}{2}+\frac{l}{2}-\frac{1}{4})}.
\end{split}
\end{equation}

When $A -\hat{k}^2>0$ the dispersion relation must be derived by matching the coefficients of the two leading-order terms in equations (\ref{29m}) and (\ref{53}) 
\begin{equation}
\begin{split}
\label{57}
&\left({\frac{\sqrt{\hat{\gamma}/S}}{\hat{k}}}\right)^{l}\frac{(\omega+l+\frac{1}{2})}{(\omega-l+\frac{1}{2})}\frac{\Gamma(-l)}{\Gamma(l)}\frac{\Gamma\left(\frac{(\omega+l)^{2}-\frac{1}{4}}{4\omega}\right)}{\Gamma\left(\frac{(\omega-l)^{2}-\frac{1}{4}}{4\omega}\right)}=\\
&\frac{\cos\left(\frac{\pi}{2}(\frac{1}{2}+l)\right)\Gamma(\frac{1}{2}-l)}{\cos\left(\frac{\pi}{2}(\frac{1}{2}-l)\right)\Gamma(\frac{1}{2}+l)}\frac{\Gamma(l)}{\Gamma(-l)}\frac{\Gamma(\frac{\nu}{2}-\frac{l}{2}+\frac{5}{4})\Gamma(\frac{\nu}{2}-\frac{l}{2}-\frac{1}{4})}{\Gamma(\frac{\nu}{2}+\frac{l}{2}+\frac{5}{4})\Gamma(\frac{\nu}{2}+\frac{l}{2}-\frac{1}{4})}\times\\
&\Bigg[{\left(1+\frac{\Gamma(1-\nu)\Gamma(\frac{\nu}{2}+\frac{l}{2}+\frac{5}{4})\Gamma(\frac{\nu}{2}+\frac{l}{2}-\frac{1}{4})}{\Gamma(1+\nu)\Gamma(-\frac{\nu}{2}+\frac{l}{2}+\frac{5}{4})\Gamma(-\frac{\nu}{2}+\frac{l}{2}-\frac{1}{4})}\right)}\Big/\\
&{\left(1+\frac{\Gamma(1-\nu)\Gamma(\frac{\nu}{2}-\frac{l}{2}+\frac{5}{4})\Gamma(\frac{\nu}{2}-\frac{l}{2}-\frac{1}{4})}{\Gamma(1+\nu)\Gamma(-\frac{\nu}{2}-\frac{l}{2}+\frac{5}{4})\Gamma(-\frac{\nu}{2}-\frac{l}{2}-\frac{1}{4})}\right)}\Bigg].
\end{split}
\end{equation}

In the limit of no stratification (i.e. $A = 0$), equation (\ref{56}) reduces to the well-known dispersion relation of the standard tearing mode instability \citep{2} while the spectral range of validity of equation (\ref{57})  narrows to zero.

\section{Results}
\label{sec6}
In this section, we compare the analytical results with the numerical solutions of equations (\ref{10}) and (\ref{12}) for both $A>0$ and $A<0$ to determine the effect of stratification on equilibrium stability. The numerical integration of equations (\ref{10}) and (\ref{12}) was performed using an adaptive finite difference scheme based on Newton iteration which was designed by Lentini and Pereyra to solve boundary value problems for systems of ordinary differential equations \citep{5}. This method was applied to compute the eigenvalues and eigenfunctions for two cases: the linear density profile (equation \ref{6}) and a more realistic profile described by a hyperbolic tangent (equation \ref{58}). For the analytical dispersion relations given in equations (\ref{56})-(\ref{57}), we employed a Newton-type root-finding method, which iteratively finds zeros of a function starting from an initial guess and converges to the nearest root. Our results are summarized in figures \ref{fig1}-\ref{fig6}.

\subsection{Effect of linear stratification (linear profile)}
\label{sec4.1}
According to the definition of the stratification parameter given in equation (\ref{parameter_A}), for the linear density profile $A$ can be expressed as
\begin{equation}
\label{linear_A}
    A = \frac{a\,g}{V_{A}^{2}}\,\epsilon.
\end{equation}
\begin{figure}
\centering
\includegraphics[width=0.8\linewidth]{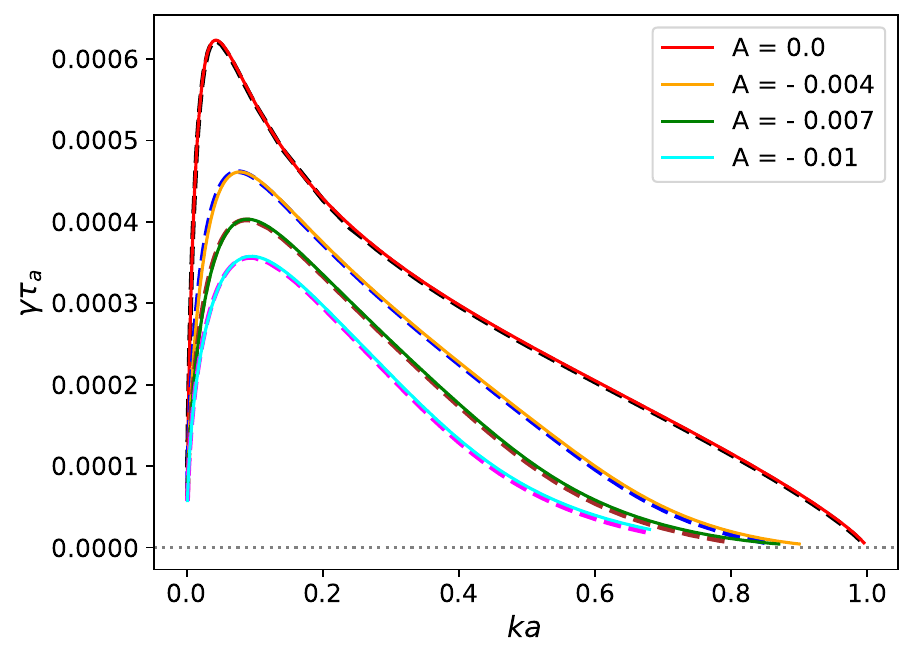}
\caption{The analytical (solid lines) and numerical (dashed lines) dispersion relation for different favorable stratification values ($A\leq0$) for $S=10^6$.}
\label{fig1}
\end{figure}
For stable stratification ($A<0$), figure \ref{fig1} shows that as $|A|$ increases, the growth rate decreases, the range of unstable modes narrows, and the fastest growing mode shifts toward smaller wavelengths. These results are consistent with \citep{3} and show excellent agreement between the analytic dispersion relation (solid lines) and the numerical solution to the eigenvalue problem (dashed lines). 

\begin{figure}
\centering
\includegraphics[width=0.8\textwidth]{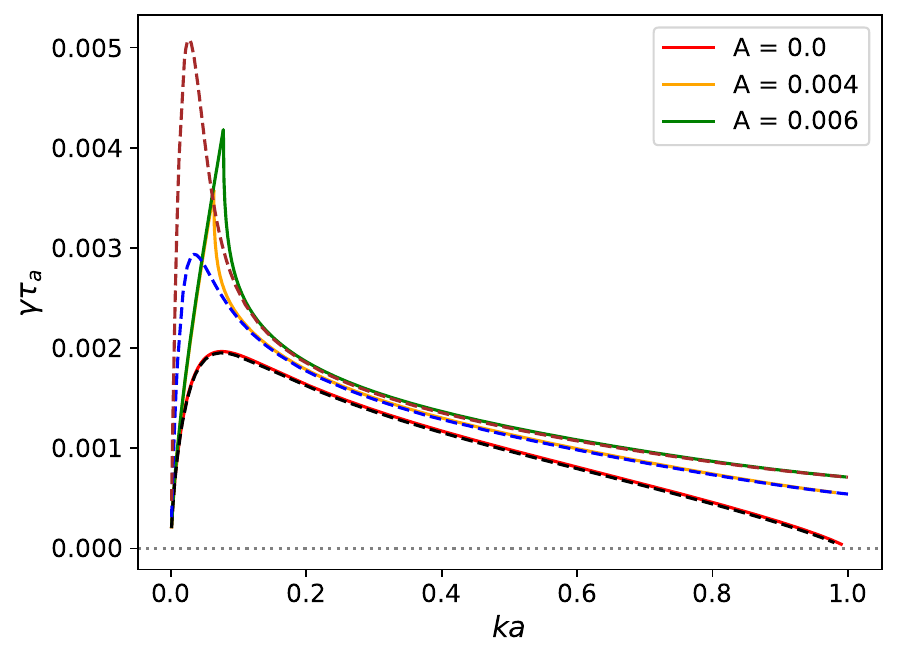}
\caption{The analytical (solid lines) and numerical (dashed lines) dispersion relation for different unfavorable stratification values ($A\geq0$) for $S=10^5$.}
\label{fig2}
\end{figure}

Conversely, figure \ref{fig2} shows that for unfavorable stratification ($A > 0$) the tearing mode growth rate increases with $A$, and unstable modes extend to values $\hat{k} > 1$. Note that in this figure the solid lines illustrate the dispersion relations (\ref{56}) and (\ref{57}), each within their valid regimes, namely $\hat{k}^{2} > A$ and $ \hat{k}^2 < A$, respectively.  
The transition from oscillatory to exponentially localized  solutions of the outer equation (\ref{15}) occurs at $\hat{k}^{2} = A$, producing peaks in the analytical results that are absent in the numerical solutions. This discrepancy arises from the finite domain used in numerical integration, which introduces a mismatch for $\hat k^{2} \lesssim A$ that is reduced by increasing the length of the integration domain. As discussed in Section \ref{sec3}, this is an unphysical behavior that can be mitigated by adopting a density profile that asymptotes to constant values far from the inner layer, ensuring spatial localization of the eigenfunction. This case will be discussed next. Despite this discrepancy, excellent agreement between numerical and analytical solution is found for $\hat k^2>A$. In this regime, stratification with $A > 0$ significantly increases the growth rate. Furthermore, unfavorable stratification  couples the tearing mode with the gravity-driven G-mode thereby eliminating the marginally stable mode. In figures \ref{fig1}-\ref{fig2}, the unstratified case is recovered for $A = 0$ (red color), where the marginally stable mode asymptotically approaches $\hat{k} \rightarrow 1$.

\subsection{Effect of stratification with smooth transition (Tanh profile)}
\label{sec4.2}

A more realistic equilibrium density profile, instead of the form presented in equation (\ref{6}), can be expressed as 
\begin{equation}
    \label{58}
        \rho_{0}(x) = \bar{\rho}_{0}\left(1+\epsilon \tanh\left( \frac{x}{a}\right)\right),
\end{equation}
which recovers the linear density profile for $|x|/a\ll1$ but introduces a finite and smooth transition region between two well-defined plasma sectors. In this case, the equilibrium state must satisfy
\begin{equation}
\label{60}
P_0(x)+\frac{B_{0}^{2}}{2\mu_{o}}+\bar{\rho}_{0}\left(x+\epsilon\,a\ln\left(\cosh{\left(\frac{x}{a}\right)}\right)\right)g = P_0(0),
\end{equation}
while the stratification parameter $A$ becomes
\begin{equation}
\label{tanh_A}
A = \frac{a\,g}{V_{A}^{2}}\,\epsilon\,\text{sech}^{2}{\left(\frac{x}{a}\right)}.
\end{equation}
Note that the parameter $A$ shown in the plots corresponds to evaluating this function at $x=0$, which gives exactly the stratification $A$ of the linear profile (equation \ref{linear_A}).

\begin{figure}
\centering
\includegraphics[width=\textwidth]{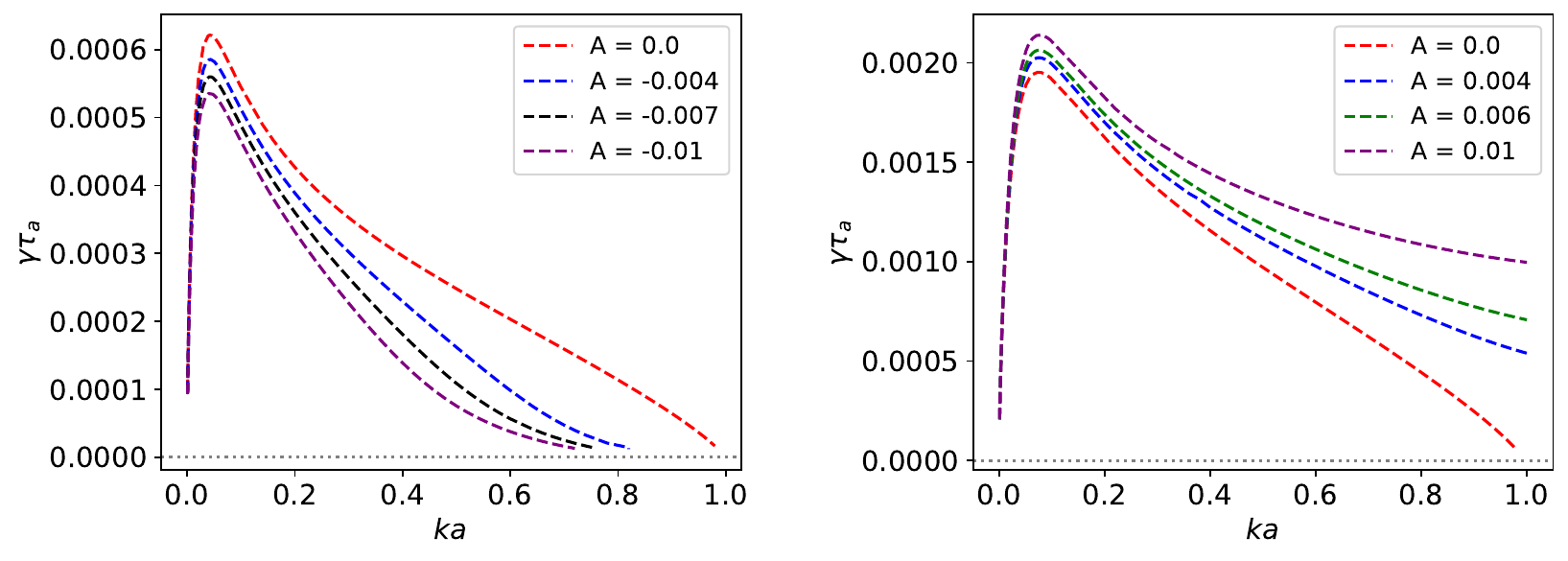}
\caption{ Dispersion relation obtained by solving the eigenvalue problem numerically for the hyperbolic tangent density profile given in equation (\ref{58}) for $A\leq0$ and $S=10^6$ (left panel) and $A\geq0$ and $S=10^5$ (right panel).}
\label{fig3}
\end{figure}
As shown in figure \ref{fig3}, the presence of both favorable and unfavorable stratification  modifies the growth rate for the $tanh$ profile mostly at large wave vectors, $\hat{k}\gtrsim \hat{k}_{m}$ where $\hat{k}_m$ is the wave vector of the fastest growing mode for $\hat{k}<1$, particularly those pertaining to the constant-$\psi$ regime ($\hat{k}\lesssim1$). On the contrary, the growth rate in the non-constant-$\psi$ regime remains essentially unaffected. More specifically,  the left panel shows that the growth rate decreases and the range of unstable modes narrows with increasing stratification strength $|A|$. On the other hand, the right panel illustrates that the system becomes less stable and the growth rate increases as the parameter $A$ increases leading to a modified tearing regime. Again, a marginal stable mode does not exist for $A>0$, indicating a smooth transition between tearing ($\hat{k}<1$) and the G-mode at values $\hat{k}>1$. 

\begin{figure}
    \centering
    \includegraphics[width=0.7\linewidth]{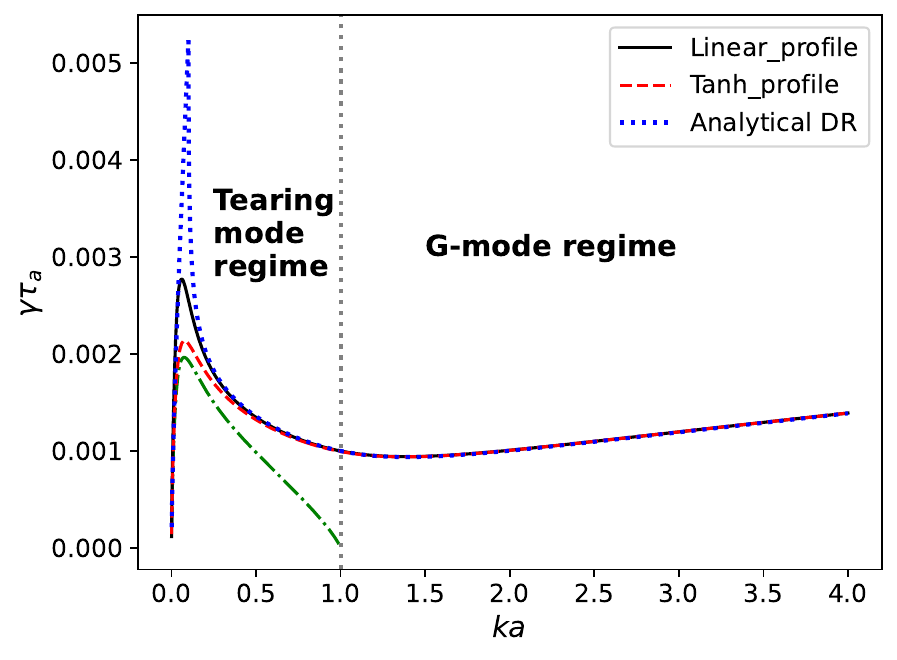}
    \caption{The analytical (dotted line) and numerical (solid and dashed lines)  dispersion relation highlighting the effect of stratification on the tearing mode and the region of instability of G-modes for the density profiles given in equations (\ref{6}) and (\ref{58}) for $S= 10^{5}$ and $A = 0.01$. The green dot-dashed line corresponds to $A=0$.}
    \label{fig4}
\end{figure}

Figure \ref{fig4} illustrates  the numerical (black solid and red dashed lines) and analytical (blue dotted line) dispersion relation in a wider range of wave vectors for $A=0.01$. For $\hat{k}<1$ the tearing mode in the constant-$\psi$ regime is strongly modified by gravity, whereas the resistive G-mode develops at wave vectors $\hat{k} \gtrsim 1 $, as discussed above. As a consequence of the transition from the tearing mode to the resistive G-mode, the growth rate remains nonzero at $\hat{k} = 1$, contrary to standard tearing theory (green dot-dashed line). Once in the G-mode regime, the growth rate increases with the wave vector and is independent of the equilibrium density distribution.  As already discussed, the effect of the equilibrium becomes more pronounced for $\hat{k}^2\simeq A$, as the tearing mode has higher values for the linear profile within that range of wave vectors. 

\begin{figure}
\centering
\includegraphics[width=\textwidth]{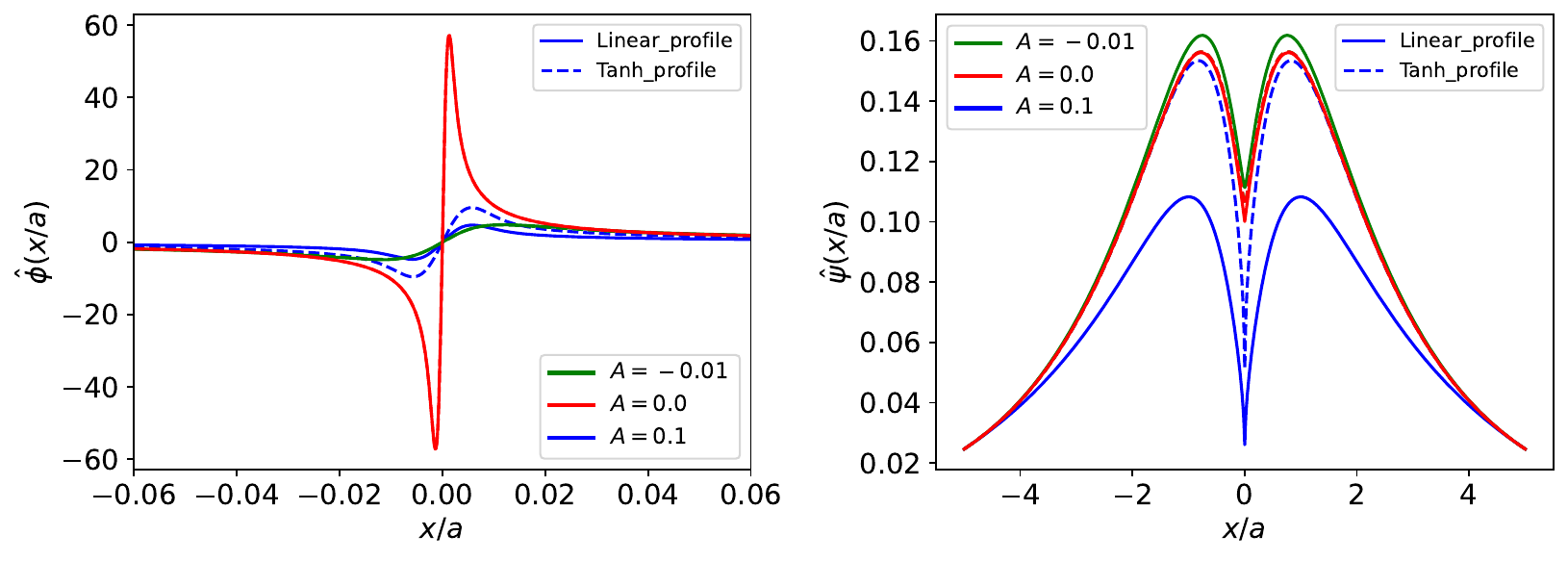}
\caption{Numerically solved normalized tearing mode eigenfunctions for $S = 10^7$ and $ka = 0.5$ for three values of $A$: velocity field stream function $\hat{\phi}$ (left panel) and magnetic field flux function $\hat{\psi}$ (right panel).}
\label{tearing}
\end{figure}

\begin{figure}
\centering
\includegraphics[width=\textwidth]{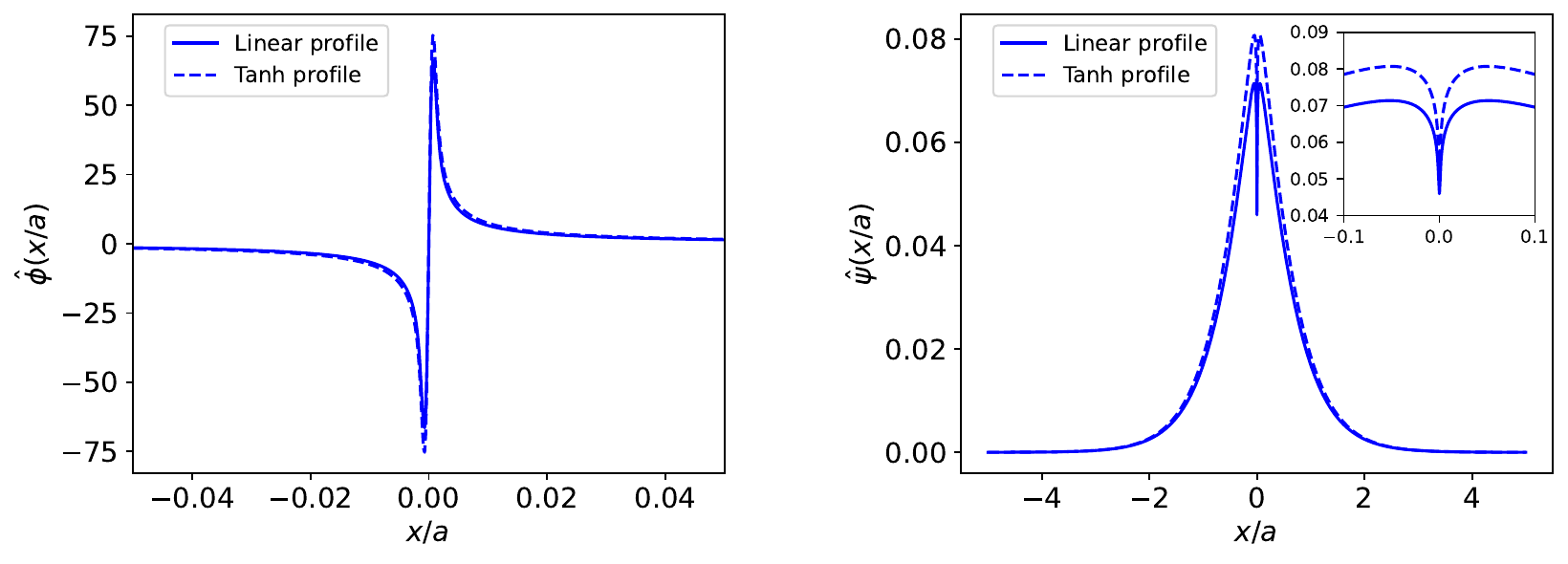}
\caption{Numerically solved normalized G-mode eigenfunctions for $S = 10^9$, $ka = 2.0$, and $A=0.1$: velocity field stream function $\hat{\phi}$ (left panel) and magnetic field flux function $\hat{\psi}$ (right panel).}
\label{Gmode}
\end{figure}

In figure \ref{tearing} we present the eigenfunctions of the tearing mode obtained by solving equations (\ref{10}) and (\ref{12}) for $S = 10^7$ and $\hat{k} = 0.5$, considering zero, favorable, and unfavorable stratification regimes. The shapes of the eigenfunctions do not differ considerably across the three regimes; however, the amplitudes of $\hat{\psi}$ and $\hat{\phi}$ change with the parameter $A$. Particularly, the amplitude of flux function $\hat{\psi}$ increases as $A$ decreases, while the stream function $\hat{\phi}$ attains its largest amplitude in the limit of no stratification. The eigenfunctions of the G-mode for $S = 10^9$, $A=0.1$, and $\hat{k} = 2.0$ are shown in figure \ref{Gmode}. It can be seen that both the flux and stream functions are very localized around the singular layer and have the same parity as the tearing mode. Finally, in both modes, the $tanh$ profile has a slightly higher amplitude than the linear profile in the $A>0$ case. 

\subsection{Scaling laws with $S$}
\label{sec4.3}

As shown in previous sections, stratification has a significant effect only at large wave vectors and has minimal influence at small wave vectors when a realistic density profile is adopted. It is also well established in classical tearing theory that the constant-$\psi$ regime is characterized by $\omega\ll1$, whereas $\omega\sim 1$ in the non-constant-$\psi$ regime. In this sense, stratification effects are negligible in the non-constant-$\psi$ regime, since $|A|\ll1$. These observations motivate us to focus on the case $\omega\ll1$ and $\hat{k}^{2}>A$, for which the analytic dispersion relation is valid for both $A\geq 0$ and $A<0$, in order to determine how the scaling of the growth rate changes with $S$ as a result of the presence of stratification. The analytical dispersion relation given in equation (\ref{56}) will admit a well-defined power law when specific asymptotic limits of $A$ and $\omega$  are considered. In what follows, we apply the simultaneous limits $\omega \ll1$ and $|A|\ll1$, along with $S\rightarrow \infty$ to only the left-hand side of equation (\ref{56}). Since the right-hand side does not depend on $S$, it will not contribute to the scaling of $\gamma\tau_a$ with $S$. A complete derivation of the scaling laws is reported in Appendix \ref{AppendixC} and the main results are summarized below.

In the regime $|A| \ll \omega\ll1$, the growth rate scales as $\gamma\tau_a \sim S^{-3/5}$, thus recovering the known constant-$\psi$ scaling law. Conversely, in the limit $\omega\ll |A|\ll1$, the scaling changes to $\gamma\tau_a \sim S^{-1}$, which means the instability becomes diffusive. As noted in Appendix \ref{AppendixC}, this regime exists only for $A<0$.  Finally,  when $A\approx \omega \ll1$ and $A>0$, the growth rate $\gamma\tau_a \sim S^{-1/3}$, recovering the resistive G-mode scaling. We therefore conclude that the inclusion of stratification bounds the scaling laws of the growth rate between two limiting values, namely $\gamma\tau_a\sim S^{-1}$ and $\gamma\tau_a\sim S^{-1/3}$. Furthermore, for a finite and fixed value of $A$ and increasing values of the Lundquist number, the constant-$\psi$ tearing mode regime will transition towards a slow diffusive mode if $A<0$ or to the G-mode, which is a rapidly growing reconnecting mode.

\begin{figure}
\centering   
\includegraphics[width=0.7\textwidth]{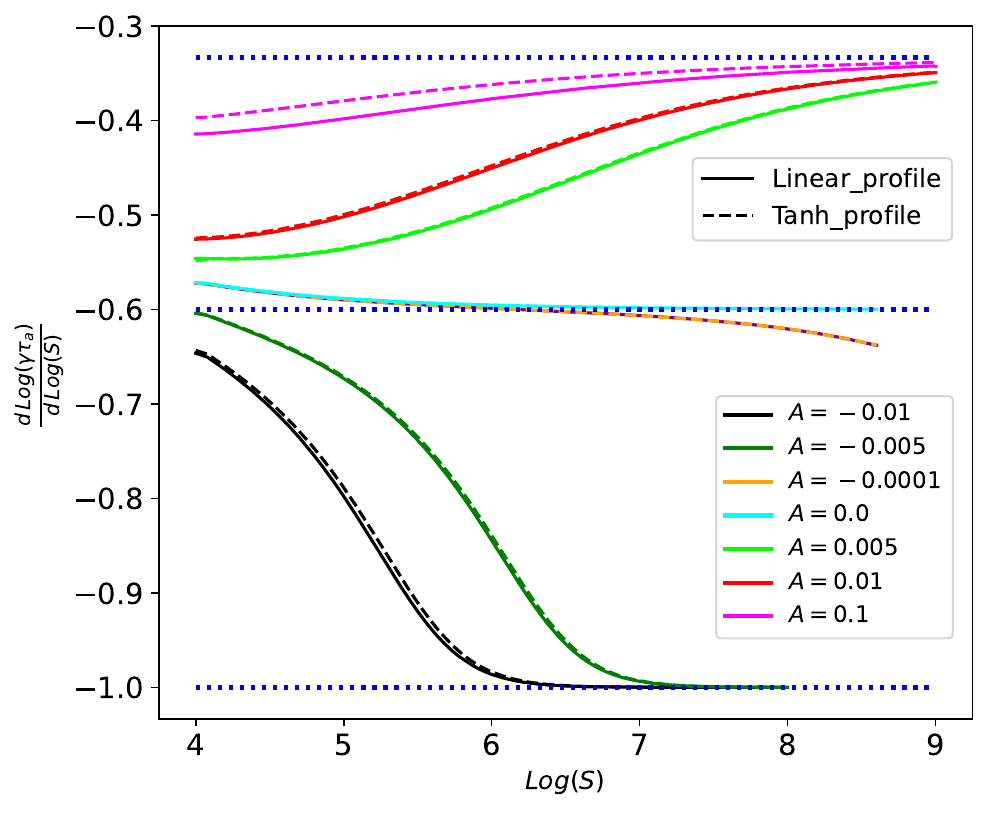}
\caption{The slope of the logarithm of $\gamma\tau_a$ as a function of the logarithm $S$, obtained from numerical solution, for different stable and unstable stratification levels and $ka = 0.5$. The blue dotted lines denote the reference scaling $d\,log(\gamma\tau_{a})/d\,log(S) = -1, -3/5, -1/3$, ordered from bottom to top. Solid lines correspond to the linear density profile and dashed lines to the hyperbolic tangent profile.}
\label{fig5}
\end{figure}
Figure \ref{fig5} presents numerically how the slope of $\log\hat\gamma$ as a function of $\log S$ varies for different stratification values. We find that the power-law scaling asymptotes to $\gamma\tau_{a}\sim S^{-1}$ if $A<0$ and $\gamma\tau_{a}\sim S^{-1/3}$ if $A>0$ as predicted, while if $A = 0$ the growth rate scales as $\gamma\tau_{a}\sim S^{-3/5}$.  Moreover, our results are in agreement with the weak and strong stable stratification regimes studied by \citep{3}. Notably, the $tanh$ density profile displays the same predicted trends.
\begin{figure}
\centering   
\includegraphics[width=0.7\textwidth]{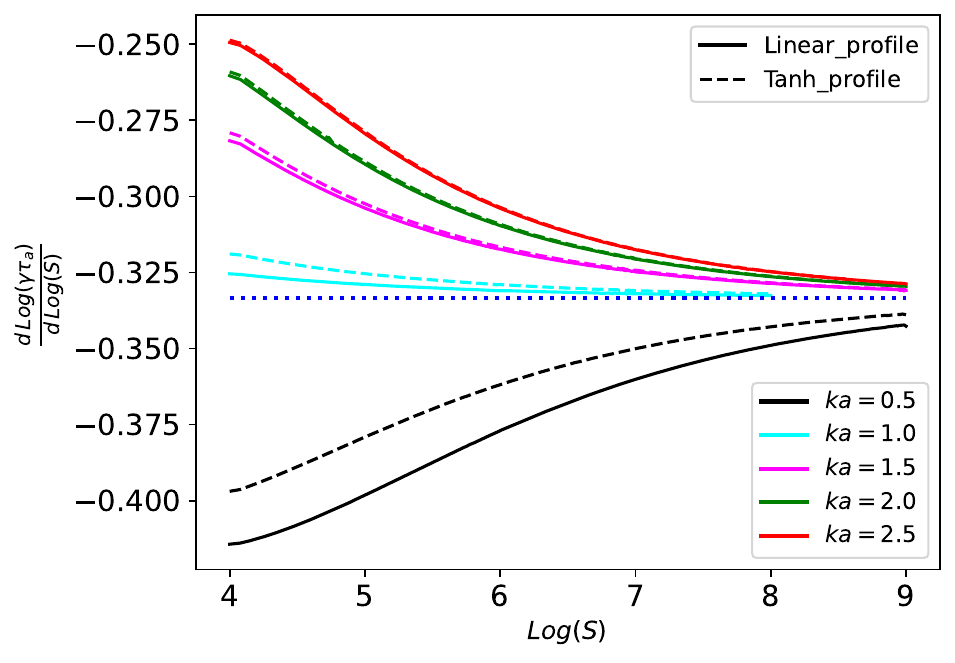}
\caption{The slope of the logarithm of $\gamma\tau_a$ as a function of the logarithm $S$, determined from numerical solution, in the large stratification limit (i.e. $A = 0.1$) for different wave vectors. The blue dotted line shows the reference scaling $d\,log(\gamma\tau_{a})/d\,log(S)=-1/3$. }
\label{fig6}
\end{figure}

In the favorable stratification asymptotic regime, the tearing mode growth rate scales as $\gamma\tau_A \sim S^{-1}$, implying that the instability grows on the same timescale as the resistive diffusion of the equilibrium current sheet, $\tau_R = S\tau_A$. This result has an important implication for the interpretation of the instability. In standard linear tearing theory, the equilibrium magnetic field is assumed to evolve resistively much more slowly than the instability growth, allowing the equilibrium to be treated as effectively fixed (quasi-static) during the linear phase. However, when $\gamma\tau_A \sim S^{-1}$, this assumption is no longer satisfied because resistive diffusion and instability growth occur at comparable rates. As a result, the equilibrium evolution can no longer be neglected, and the validity of the classical linear tearing framework becomes questionable in this asymptotic limit. This behavior suggests that favorable stratification suppresses the tearing instability so strongly that its growth becomes extremely slow, approaching a marginal regime in which the instability competes directly with resistive diffusion of the current sheet. Therefore, a more self-consistent treatment that includes the resistive evolution of the equilibrium may be required to accurately describe the system dynamics.

It is also interesting to compare the growth rate scaling with $S$ for the tearing and G-modes in the unstable stratification limit ($A>0$), shown in figure \ref{fig6} for $A=0.1$ and different wave vectors. It can be noticed that as $S$ increases, the power law of the growth rate for a given wave vector becomes closer and closer to the value $-1/3$ in both the linear and $tanh$ density profile. Specifically, for the tearing mode ($\hat{k}\lesssim 1$), the stratification significantly alters the scaling of the constant-$\psi$ regime, gradually changing it from $\gamma\tau_{a} \sim S^{-3/5}$ to $\gamma\tau_{a} \sim S^{-1/3}$. The same asymptotic power law of $-1/3$ is reached for larger wave vectors ($\hat{k}>1$), consistent with the well established scaling law of the G-mode \citep{6}. This indicates that for a fixed value of $A>0$, the gravity-modified tearing mode progressively transitions to the G-mode as $S$ increases. In other words, the constant-$\psi$ regime effectively does not exist, even for weak unfavorable stratification, when $S\gg1$. A useful estimate for the onset of stratification effects can be obtained from the crossover condition ($|A| \sim \omega$), which separates the weak-stratification regime ($|A| \ll \omega$), for which the classical constant-$\psi$ tearing scaling is recovered, from the stratification-modified regime discussed in Appendix \ref{AppendixC}. Using the classical constant-$\psi$ tearing scaling, $\gamma\tau_a\sim S^{-3/5}(\Delta' a)^{4/5}(ka)^{2/5}$, we obtain the following estimate for the crossover value of the stratification parameter:
\begin{equation*}
A_{\mathrm{crit}}\sim S^{-2/5}(ka)^{-2/5}(\Delta'a)^{6/5}.
\end{equation*}
This result implies that the threshold decreases with increasing Lundquist number, explaining why even weak stratification can significantly modify the tearing growth rate at sufficiently large (S), as illustrated in figure \ref{fig5}.

\section{Summary and conclusions}
\label{sec7}
In this paper, we carried out a linear stability analysis of a slab current sheet by including a gravitational acceleration acting normal to the current layer, and studied how density stratification and gravity influence the equilibrium stability in both favorable and unfavorable density configurations. We first derived the  linearized equations for the perturbed velocity and magnetic field assuming a constant density gradient, and obtained the dispersion relation through asymptotic matching of the outer and inner solution.  These analytical predictions were then compared with numerical solutions of the full eigenvalue problem. To determine how the instability depends on the assumed density model, we also introduced a more realistic hyperbolic tangent density profile and compared the resulting growth rate with the theoretical predictions. Our main findings are summarized below. 

Favorable stratification ($A<0$) suppresses perturbations and enhances the stability of the tearing mode. As $|A|$ increases, the growth rate decreases, the range of unstable modes narrows, and the fastest growing mode shifts toward smaller wavelengths.

Unfavorable stratification ($A>0$) qualitatively changes the structure of the outer solution: solutions transition from oscillatory behavior for $\hat{k}^{2} < A$ to exponential localization as $|x| \to \infty$ for $\hat{k}^{2} > A$. The growth rate increases with $A$ and the unstable spectrum extends to the G-mode regime at $\hat{k}>1$. We found a mismatch for $\hat{k}^2\lesssim A$ between analytical and numerical results for the linear density profile, due to the unlimited extent of the constant density gradient. This inconsistency is removed when using a smooth hyperbolic tangent profile, which approaches constant density away from the current layer. 

For the $tanh$ density profile, the dependence of the growth rate on stratification closely matches that obtained with the linear profile. Favorable stratification again reduces the growth rate and narrows the unstable spectrum. In the unfavorable case, the dispersion relation remains regular across $\hat{k}^{2} = A$, and stratification primarily affects the large‑$k$ behavior. For $\hat{k}^{2} > A$, numerical results agree well with analytical predictions: stratification causes a gravity‑modified tearing regime for $\hat{k} \lesssim 1$ that transitions smoothly into the resistive G‑mode. This transition occurs once gravitational and stratification effects dominate, destabilizing the resistive G‑mode for $\hat{k} > 1$. In both favorable and unfavorable regimes, the maximum growth rate and corresponding wave vector show only weak dependence on the stratification strength.

To understand the transition from tearing to G-modes, we investigated the growth rate scaling with $S$ for different stratification regimes by using the analytical dispersion relation, and compared predictions with numerical results.  We found that the known constant-$\psi$ scaling $\gamma\tau_{a}\sim S^{-3/5}$ is recovered provided stratification becomes weaker for increasing $S$ ($A\ll\sqrt{\hat\gamma^3 S/k^{2}}\ll 1$). For a finite stratification parameter $A$, however, the growth rate admits two asymptotic limits that provide a lower and upper bound to the growth rate, namely $\gamma\tau_{a}\sim S^{-1}$ for $A<0$  and $\gamma\tau_{a}\sim S^{-1/3}$ for $A>0$. 

In conclusion, a favorable stratification suppresses reconnection, while unfavorable stratification only allows for rapidly reconnecting modes with $\gamma\tau_{a}\sim S^{-1/3}$, namely, the gravity-driven G-mode and the non-constant-$\psi$. In this work we have neglected  physical effects that become non negligible when dispersive or kinetic scales are reached, possibly modifying the dispersion relation at large wave vectors, or when considering microscopic current sheets. Investigating these extensions will be the subject of future work.

\appendix

\section{Solution of the outer equation}
\label{AppendixA}
The general solution of equation (\ref{15}) in the vicinity of the boundary layer ($|\hat{x}|\rightarrow0$) is a linear combination of two power series determined via Frobenius expansion as follows 
\begin{equation}
\label{16}
    \hat{\phi}(\hat{x}) = c_{1}\hat{x}^{-l-\frac{1}{2}} +c_{2} \hat{x}^{l-\frac{1}{2}},
\end{equation}
where $l = \sqrt{\frac{1}{4}-A}$ and $c_{1}$ and $c_{2}$ are two constants to be determined by matching with the limit $|\hat x|\rightarrow0$ of the complete solution. Following the method of \citep{3} in order to get the complete solution of the outer equation (\ref{15}) we apply the transformation $T = \tanh^2(\hat{x})$, leading to the following equation,
\begin{equation}
    \label{17}
    \frac{d^{2}\hat{\phi}}{dT^{2}}+\left(\frac{3/2}{T(1-T)}-\frac{5/2}{(1-T)}\right)\frac{d\hat{\phi}}{dT}-\frac{\nu^{2}T +(l^{2}-\frac{1}{4})(1-T)}{4T^{2}(1-T)^{2}}\hat{\phi} =0,
\end{equation}
with $\nu = \sqrt{\hat{k}^{2}-A}$. Since this equation has three regular singular points at $0, 1,$ and $\infty$, it can be recast into the standard form of the hypergeometric differential equation through an appropriate change of variables as follows
\begin{equation}
    \label{18}
\hat{\phi}(T) = T^{p} (1-T)^{q}W(T).
\end{equation}
With some algebra, we obtain 
\begin{equation}
    \label{19}
    T(1-T)W^{''}+\left[(2p+\frac{3}{2})-(2p+2q+\frac{5}{2})T\right]W^{'}-\left[(p+q)^{2}+\frac{3}{2}(p+q)\right]W=0,
\end{equation}
which possesses the following constraints 
\begin{align}
    \label{20}
 q^{2}- \frac{\nu^{2}}{4} &= 0;& p^{2}+\frac{p}{2}-\frac{1}{4}\left(l^{2}-\frac{1}{4}\right) = 0. 
\end{align}
The solution of equation (\ref{15}) is sought near the singular point $T=1$ ($\hat{x}=\infty$) to ensure it is well behaved or convergent as $\hat{x}\rightarrow\infty$. Near this siguglar point the solution takes the form 
\begin{align}
        \label{21}
    \hat{\phi}\left(\hat{x}\right)&= T^{\frac{l}{2}-\frac{1}{4}} \left(1-T\right)^{\frac{\nu}{2}}\,_{2}F_{1}\left(\frac{\nu}{2}+\frac{l}{2}-\frac{1}{4}, \frac{\nu}{2}+\frac{l}{2}+\frac{5}{4}, 1+\nu,1-T\right)\nonumber\\&+\alpha_{2}T^{\frac{l}{2}-\frac{1}{4}}\left(1-T\right)^{-\frac{\nu}{2}}\,_{2}F_{1}\left(\frac{l}{2}-\frac{\nu}{2}-\frac{1}{4},\frac{l}{2}-\frac{\nu}{2}+\frac{5}{4},1-\nu,1-T\right)\,\,\mathrm{sgn}(\hat{x}),
\end{align}
where $\alpha_{2}$ is an arbitrary constant and $\,\mathrm{sgn}(\hat{x})$ ensures antisymmetry. Given that the parameter $\nu$ may be real or imaginary depending on the values of $A$ and $\hat{k}^2$, it is necessary to analyze the asymptotic behavior  of the obtained solution (\ref{21}) as $|\hat{x}| \rightarrow \infty$ separately for each case: if $\hat{k}^2> A$ , then $\nu$ is purely real and only the first term of equation (\ref{21}) is retained, as it is well behaved in the limit $|\hat{x}| \rightarrow \infty$,
\begin{equation}
        \label{22}
    \hat{\phi}(\hat{x})=  T^{\frac{l}{2}-\frac{1}{4}}\left(1-T\right)^{\frac{\nu}{2}}\,_2F_1 \left(\frac{l}{2}+\frac{\nu}{2}-\frac{1}{4},\frac{l}{2}+\frac{\nu}{2}+\frac{5}{4},1+\nu,1-T\right)\,\,\mathrm{sgn}(\hat{x}).
\end{equation}
On the other hand, when $\hat{k}^2< A$, the parameter $\nu$ becomes purely imaginary. But the flow function is a real-valued function, and hence its Fourier coefficients must satisfy Hermitian symmetry (i.e. $\hat{\phi}^{*}(\hat{k}) = \hat{\phi}(-\hat{k})$). As a result, we must retain both terms in equation (\ref{21}), with $\alpha_{2}$ fixed to one, to capture the full behavior of the solution:  
\begin{align}
        \label{23}
    \hat{\phi}(\hat{x})&= \Bigg[T^{\frac{l}{2}-\frac{1}{4}} (1-T)^{\frac{\nu}{2}}\,_{2}F_{1}\left(\frac{\nu}{2}+\frac{l}{2}-\frac{1}{4}, \frac{\nu}{2}+\frac{l}{2}+\frac{5}{4}, 1+\nu,1-T\right)\nonumber\\&+T^{\frac{l}{2}-\frac{1}{4}} (1-T)^{-\frac{\nu}{2}}\,_{2}F_{1}\left(\frac{l}{2}-\frac{\nu}{2}-\frac{1}{4},\frac{l}{2}-\frac{\nu}{2}+\frac{5}{4},1-\nu,1-T\right)\Bigg]\,\mathrm{sgn}(\hat{x}).
\end{align}
We now make use of the following connection formula for the hypergeometric function to relate the solution around $T=1$ ($\hat x=\infty$) to the solution around $T=0$ ($\hat x=0$), 
\begin{align}
    \label{24}
    \,_2F_1(a,b;c;z)=& \frac{\Gamma(c)\Gamma(c-a-b)}{\Gamma(c-a)\Gamma(c-b)}\,_2F_1(a,b;a+b+1-c;1-z)\nonumber\\&+\frac{\Gamma(c)\Gamma(a+b-c)}{\Gamma(a)\Gamma(b)}(1-z)^{c-a-b}\,_2F_1(c-a,c-b;1+c-a-b;1-z).
\end{align} 
Finally, with the following series expansion \citep{abramowitz1965handbook}, 
\begin{equation}
    \label{25}
        \,_2F_1(a,b;c;z)= \sum_{n=0}^{n=\infty}\frac{(a)_{n} (b)_{n}}{(c)_{n}}\frac{z^{n}}{n!} = \frac{\Gamma(c)}{\Gamma(a)\Gamma(b)} \sum_{n=0}^{n=\infty}\frac{\Gamma(a+n)\Gamma(b+n)}{\Gamma(c+n)} \frac{z^{n}}{n!},
\end{equation}
the outer layer solution (\ref{22}) can be written in the following series form 
\begin{equation}
        \label{26}
    \hat{\phi}(\hat{x})= \sum_{n=0}^{\infty}\left[a_n |T|^{n-\frac{l}{2}-\frac{1}{4}}+b_n |T|^{n+\frac{l}{2}-\frac{1}{4}}\right]\left(1-T\right)^{\frac{\nu}{2}}\,\mathrm{sgn}(\hat{x}),
\end{equation}
where the coefficients $a_n$ and $b_n$ are given by 
\begin{equation}
\begin{split}
\label{27}
a_{n} = &\frac{\Gamma(1+\nu)\Gamma(l)}{\Gamma(\frac{\nu}{2}+\frac{l}{2}+\frac{5}{4})\Gamma(\frac{\nu}{2}+\frac{l}{2}-\frac{1}{4})}\frac{\Gamma(1-l)}{\Gamma(\frac{\nu}{2}-\frac{l}{2}+\frac{5}{4})\Gamma(\frac{\nu}{2}-\frac{l}{2}-\frac{1}{4})}\times\\
&\frac{\Gamma(\frac{\nu}{2}-\frac{l}{2}+\frac{5}{4}+n)\Gamma(\frac{\nu}{2}-\frac{l}{2}-\frac{1}{4}+n)}{\Gamma(1-l+n)n!},
\end{split}
\end{equation}

\begin{equation}
\begin{split}
 \label{28}
b_{n} = &\frac{\Gamma(1+\nu)\Gamma(-l)}{\Gamma(\frac{\nu}{2}-\frac{l}{2}+\frac{5}{4})\Gamma(\frac{\nu}{2}-\frac{l}{2}-\frac{1}{4})}\frac{\Gamma(l+1)}{\Gamma(\frac{\nu}{2}+\frac{l}{2}+\frac{5}{4})\Gamma(\frac{\nu}{2}+\frac{l}{2}-\frac{1}{4})}\times \\ &\frac{\Gamma(\frac{\nu}{2}+\frac{l}{2}+\frac{5}{4}+n)\Gamma(\frac{\nu}{2}+\frac{l}{2}-\frac{1}{4}+n)}{\Gamma(l+1+n)n!}.
\end{split}
\end{equation}
In the immediate vicinity of the inner layer,  this solution reduces to 
\begin{equation}
    \label{29a}
    \hat{\phi}(\hat{x})= \left[a_{0} |\hat{x}|^{-l-\frac{1}{2}}+b_{0} |\hat{x}|^{l-\frac{1}{2}}\right]\,\mathrm{sgn}(\hat{x}),
\end{equation}
with
\begin{equation}
    \label{30a}
     a_{0} = \frac{\Gamma(1+\nu)\Gamma(l)}{\Gamma(\frac{\nu}{2}+\frac{l}{2}+\frac{5}{4})\Gamma(\frac{\nu}{2}+\frac{l}{2}-\frac{1}{4})},
\end{equation}
\begin{equation}
    \label{31a}
     b_{0} = \frac{\Gamma(1+\nu)\Gamma(-l)}{\Gamma(\frac{\nu}{2}-\frac{l}{2}+\frac{5}{4})\Gamma(\frac{\nu}{2}-\frac{l}{2}-\frac{1}{4})}.
\end{equation}
By comparing equation (\ref{29a}), with equations (\ref{30a}) and (\ref{31a}), with the one obtained from the Frobenius expansion, equation (\ref{16}), enables us to determine the constants $c_{1}$ and $c_{2}$, namely $c_1={a_{0}}$ and $c_{2}=b_{0}$. Keep in mind that when $\hat{k}^2 < A$, these coefficients are modified due to the contribution of the second term that was dropped for $\hat{k}^2 > A$. Repeating the same steps for the $\hat{k}^2 < A$ case leads to the following outer solution
\begin{equation}
    \label{29ma}
    \hat{\phi}(\hat{x})= \left[a_{0mod} |\hat{x}|^{-l-\frac{1}{2}}+b_{0mod} |\hat{x}|^{l-\frac{1}{2}}\right]\,\mathrm{sgn}(\hat{x}),
\end{equation}
where the coefficients $a_{0mod}$ and $b_{0mod}$ take the form
\begin{equation}
    \label{32a}
     a_{0 mod} = \frac{\Gamma(1+\nu)\Gamma(l)}{\Gamma(\frac{\nu}{2}+\frac{l}{2}+\frac{5}{4})\Gamma(\frac{\nu}{2}+\frac{l}{2}-\frac{1}{4})}+\frac{\Gamma(1-\nu)\Gamma(l)}{\Gamma(-\frac{\nu}{2}+\frac{l}{2}+\frac{5}{4})\Gamma(-\frac{\nu}{2}+\frac{l}{2}-\frac{1}{4})},
\end{equation}
\begin{equation}
    \label{33a}
     b_{0mod} = \frac{\Gamma(1+\nu)\Gamma(-l)}{\Gamma(\frac{\nu}{2}-\frac{l}{2}+\frac{5}{4})\Gamma(\frac{\nu}{2}-\frac{l}{2}-\frac{1}{4})}+\frac{\Gamma(1-\nu)\Gamma(-l)}{\Gamma(-\frac{\nu}{2}-\frac{l}{2}+\frac{5}{4})\Gamma(-\frac{\nu}{2}-\frac{l}{2}-\frac{1}{4})}.
\end{equation}
\section{Solution of the inner equation}
\label{AppendixB}
We first introduce the Fourier transform in the following form 
\begin{equation}
\label{36}
    \hat{\psi} (\hat{x}) = \frac{1}{2 \pi} \int_{-\infty}^{\infty} \tilde{\psi}(\hat{\theta}) e^{i \hat{\theta}\hat{x}} d\hat{\theta}.
\end{equation}
\begin{equation}
\label{37}
    \hat{\phi} (\hat{x}) = \frac{1}{2 \pi} \int_{-\infty}^{\infty} \tilde{\phi}(\hat{\theta}) e^{i \hat{\theta}\hat{x}} d\hat{\theta}.
\end{equation}
Upon applying this transform, the differential equations (\ref{34})-(\ref{35}) take the following form 
\begin{equation}
    \label{38}
    \hat{\gamma} S \left(\tilde{\psi} - i \frac{\partial}{\partial \hat{\theta}} \tilde{\phi}\right) = - \hat{\theta}^{2} \tilde{\psi},
\end{equation}
\begin{equation}
      \label{39}
      \left(\hat{\theta}^{2} - A \frac{\hat{k}^{2}}{\hat{\gamma}^{2}}\right) \tilde{\phi} = - i  \frac{\hat{k}^{2}}{\hat{\gamma}^{2}} \frac{\partial}{\partial \hat{\theta}} (\hat{\theta}^{2} \tilde{\psi}),
  \end{equation}
and inserting equation (\ref{38}) into equation (\ref{39}) yields 
\begin{equation}
    \label{40}
\frac{\hat{k}^{2}}{\hat{\gamma}} S  \frac{d}{d\hat{\theta}}\left(\frac{\hat{\theta}^{2}}{\hat{\gamma}S + \hat{\theta}^{2}} \frac{d\tilde{\phi}}{d\hat{\theta}}\right)-\left(\hat{\theta}^{2} - A \frac{\hat{k}^{2}}{\hat{\gamma}^{2}}\right) \tilde{\phi} = 0.
\end{equation}
The general solution of this equation near the edge of the boundary layer ($|\hat{\theta}|\rightarrow0$) is a linear combination of two power series determined via Frobenius expansion and is given by 
\begin{equation}
    \label{41}
    \tilde{\phi}(\hat{\theta}) =(\hat{k}S)^{-\frac{l}{3}+\frac{1}{6}}d_{1}\hat{\theta}^{l-\frac{1}{2}}+(\hat{k}S)^{\frac{l}{3}+\frac{1}{6}}d_{2}\hat{\theta}^{-l-\frac{1}{2}},
\end{equation} 
where $d_{1}$ and $d_{2}$ are two constants to be determined by matching with the limit $|\hat{\theta}|\rightarrow0$ of the complete inner solution. Following the method of \citep{3,4}, in order to obtain the complete solution of the inner equation (\ref{40}) we introduce a new independent variable $\chi =\sqrt{\frac{\hat{\gamma}/S}{\hat{k}^2}}\hat{\theta}^{2}$ which yields the following equation
\begin{equation}
    \label{42}
    4\chi^{1/2}\frac{d}{d\chi}\left(\frac{\chi^{3/2}}{\chi+\omega}\frac{d\tilde{\phi}}{d\chi}\right)- \left(\chi+\frac{l^{2}-\frac{1}{4}}{\omega}\right)\tilde{\phi}=0,
\end{equation}
where $\omega = \sqrt{\frac{\hat{\gamma}^{3}S}{\hat{k}^{2}}}$. 

Applying the following transformation 
\begin{equation}
\label{43}
\tilde{\phi}(\chi) = e^{-\frac{\chi}{2}}\chi^{\frac{l}{2}-\frac{1}{4}} Z(\chi),
\end{equation}
to equation (\ref{42}) gives 
\begin{equation}
\label{44}
Z''(\chi) + \left(\frac{1+l}{\chi}-1-\frac{1}{\chi+\omega}\right)Z'(\chi)- \frac{\left(l+\omega-\frac{1}{2}\right)}{2\chi} \left[\frac{l+\omega+\frac{1}{2}}{2\omega}+\frac{1}{\chi+\omega}\right]Z(\chi) =0,
\end{equation}
where the prime denotes differentiation with respect to $\chi$. Shanin and Craster \citep{4} have provided solutions for this class of ordinary differential equations. Accordingly, following their approach, the general solution of equation (\ref{42}) can thus be expressed in terms of the confluent hypergeometric function $U(a,b,z)$ as 
\begin{equation}
\begin{split}
    \label{45}
    \tilde{\phi} (\chi)= e^{-\frac{\chi}{2}} \chi^{\frac{l}{2}-\frac{1}{4}} &\Bigg[U\left(\frac{(l+\omega)^{2}-\frac{1}{4}}{4\omega},1+l,\chi\right)\\&
+\left(\frac{(\omega-\frac{1}{2})^{2}-l^{2}}{4\omega}\right)U\left(\frac{(l+\omega)^{2}-\frac{1}{4}}{4\omega}+1,1+l,\chi\right)\Bigg]\,\mathrm{sgn}(\hat{\theta}).
\end{split}
\end{equation} 
Using Kummer's transformation for the Tricomi function $U(a,b,z)$ and noting that Kummer's function $M(a,b,z)$ in fact is a generalized hypergeometric series of the form
\begin{equation}
\label{46}
    U(a,b,z) = \frac{\Gamma(1-b)}{\Gamma(a-b+1)}M(a,b,z)+\frac{\Gamma(b-1)}{\Gamma(a)}z^{1-b}M(a-b+1,2-b,z),
\end{equation}
\begin{equation}
    \label{47}
    M(a,b,z) =\sum_{n=0}^{\infty}\frac{(a)_{n}}{(b)_{n}n!}z^{n} = 1 + \frac{a}{b}z+\frac{a(a+1)}{b(b+1)2!}z^{2}+....,
\end{equation}
then the series representation of the solution (\ref{45}) in the limit $\hat{\theta}\rightarrow 0$ can be expressed as  
\begin{equation}
    \label{48a}
      \tilde{\phi}(\hat{\theta})= \left[\tilde{a}_{0}|\hat{\theta}|^{l-\frac{1}{2}}+\tilde{b}_{0}|\hat{\theta}|^{-l-\frac{1}{2}}\right]\,\mathrm{sgn}(\hat{\theta}),
\end{equation}
where the coefficients $\tilde{a}_{0}$ and $\tilde{b}_{0}$ are
\begin{equation}
    \label{49a}
    \tilde{a}_{0}=\left({\frac{\sqrt{\hat{\gamma}/S}}{\hat{k}}}\right)^{\frac{l}{2}-\frac{1}{4}}\frac{2\omega}{(\omega-l+\frac{1}{2})}\frac{\Gamma(-l)}{\Gamma\left(\frac{(\omega-l)^{2}-\frac{1}{4}}{4\omega}\right)}.
\end{equation}
\begin{equation}
    \label{50a}
    \tilde{b}_{0}=\left({\frac{\sqrt{\hat{\gamma}/S}}{\hat{k}}}\right)^{-\frac{l}{2}-\frac{1}{4}}\frac{2\omega}{(\omega+l+\frac{1}{2})}\frac{\Gamma(l)}{\Gamma\left(\frac{(l+\omega)^{2}-\frac{1}{4}}{4\omega}\right)},
\end{equation}
By comparing equation (\ref{48a}), with equations (\ref{49a})-(\ref{50a}), with the solution given by the Frobenius expansion, equation (\ref{41}), allows us to determine the constants $d_{1}$ and $d_{2}$ as follows, 
\begin{equation}
    \label{51}
      d_{1} = (\hat{k}S)^{\frac{l}{3}-\frac{1}{6}}\left({\frac{\sqrt{\hat{\gamma}/S}}{\hat{k}}}\right)^{\frac{l}{2}-\frac{1}{4}}\frac{2\omega}{(\omega-l+\frac{1}{2})}\frac{\Gamma(-l)}{\Gamma\left(\frac{(\omega-l)^{2}-\frac{1}{4}}{4\omega}\right)},
\end{equation}
\begin{equation}
    \label{52}
      d_{2} =(\hat{k}S)^{-\frac{l}{3}-\frac{1}{6}}\left({\frac{\sqrt{\hat{\gamma}/S}}{\hat{k}}}\right)^{-\frac{l}{2}-\frac{1}{4}}\frac{2\omega}{(\omega+l+\frac{1}{2})}\frac{\Gamma(l)}{\Gamma\left(\frac{(l+\omega)^{2}-\frac{1}{4}}{4\omega}\right)}.
\end{equation}
\section{Derivation of scaling laws}
\label{AppendixC}
Let us rewrite the dispersion relation that is valid when $\hat{k}^{2}>A$ as follows
\begin{equation}
\begin{split}
\label{dr}
&\left({\frac{\sqrt{\hat{\gamma}/S}}{\hat{k}}}\right)^{l}\frac{(\omega+l+\frac{1}{2})}{(\omega-l+\frac{1}{2})}\frac{\Gamma(-l)}{\Gamma(l)}\frac{\Gamma\left(\frac{(\omega+l)^{2}-\frac{1}{4}}{4\omega}\right)}{\Gamma\left(\frac{(\omega-l)^{2}-\frac{1}{4}}{4\omega}\right)}=\\
&\frac{\cos\left(\frac{\pi}{2}(\frac{1}{2}+l)\right)\Gamma(\frac{1}{2}-l)}{\cos\left(\frac{\pi}{2}(\frac{1}{2}-l)\right)\Gamma(\frac{1}{2}+l)}\frac{\Gamma(l)}{\Gamma(-l)}\frac{\Gamma(\frac{\nu}{2}-\frac{l}{2}+\frac{5}{4})\Gamma(\frac{\nu}{2}-\frac{l}{2}-\frac{1}{4})}{\Gamma(\frac{\nu}{2}+\frac{l}{2}+\frac{5}{4})\Gamma(\frac{\nu}{2}+\frac{l}{2}-\frac{1}{4})}.
\end{split}
\end{equation}
In the small but finite stratification limit ($|A|\ll1$), the parameter $l$ can be approximated as $l \approx\frac{1}{2}-A$. Accordingly, the left-hand side simplifies to
\begin{equation}
\label{r.h.s}
\left({\frac{\hat{\gamma}}{S\hat{k}^{2}}}\right)^{\frac{1}{4}-{\frac{A}{2}}}\frac{{\omega}-A+1}{{\omega}+A}\frac{\Gamma(-\frac{1}{2}+A)}{\Gamma(\frac{1}{2}-A)}\frac{\Gamma\left(\frac{\omega^2-2\omega A +A^{2}+\omega-A}{4{\omega}}\right)}{\Gamma\left(\frac{\omega^2+2\omega A +A^{2}-\omega-A}{4{\omega}}\right)},
\end{equation}
As we mentioned above, we want to derive the scaling laws of the growth rate with Lundquist number $S$ in the regimes $|A|\ll \omega\ll1$, $\omega\ll |A|\ll1$, and $|A|\approx \omega\ll1$. As a result of those limits, the previous expression can be approximated by 
\begin{equation}
\label{sim_dr}
\left({\frac{\hat{\gamma}}{S\hat{k}^{2}}}\right)^{\frac{1}{4}}\frac{1}{{\omega}+A}\frac{\Gamma(-\frac{1}{2}+A)}{\Gamma(\frac{1}{2}-A)}\frac{\Gamma(\frac{1}{4}-\frac{A}{4\omega})}{\Gamma(-\frac{1}{4}-\frac{A}{4\omega})},
\end{equation}
where the fraction 
\begin{equation}
    \label{condition}
\frac{\Gamma(-\frac{1}{2}+A)}{\Gamma(\frac{1}{2}-A)} <0,
\end{equation}
for both positive and negative $A$, and therefore the left-hand side carries an overall negative sign. This negative sign cancels the sign arising from the subsequent scaling.
\subsection{$|A| \ll \omega\ll 1$ limit}
In the limit $|A| \ll \omega\ll 1$, we can neglect $A$ compared to $\omega$. Thus, equation (\ref{sim_dr}) reduces to  
\begin{equation}
    \label{limit1}
\left({\frac{\hat{\gamma}}{S\hat{k}^{2}}}\right)^{\frac{1}{4}}\frac{1}{{\omega}}\frac{\Gamma(-\frac{1}{2}+A)}{\Gamma(\frac{1}{2}-A)}\frac{\Gamma(\frac{1}{4})}{\Gamma(-\frac{1}{4})},
\end{equation}
and, hence, the growth rate scales with $S$ as 
\begin{equation}
\label{scale1}
\hat{\gamma} \sim S^{-3/5}.
\end{equation}

\subsection{$\omega \ll |A| \ll 1$ limit}
In the limit $\omega \ll |A| \ll 1$, we can neglect $\omega$ compared to $A$. Therefore, expression (\ref{sim_dr}) takes the form
\begin{equation}
    \label{limit2}
\left({\frac{\hat{\gamma}}{S\hat{k}^{2}}}\right)^{\frac{1}{4}}\frac{1}{{A}}\frac{\Gamma(-\frac{1}{2}+A)}{\Gamma(\frac{1}{2}-A)}\frac{\Gamma(\frac{1}{4}-\frac{A}{4\omega})}{\Gamma(-\frac{1}{4}-\frac{A}{4\omega})}.
\end{equation}
By applying the following asymptotic expansion of the Gamma function 
\begin{equation}
    \label{asympt}
    \frac{\Gamma(z+a)}{ \Gamma(z+b)}\thicksim z^{a-b},
\end{equation}
where $z=\frac{A}{4\omega}$, equation (\ref{limit2}) simplifies to
\begin{equation}
\left({\frac{\hat{\gamma}}{S\hat{k}^{2}}}\right)^{\frac{1}{4}}\frac{1}{{A}}\frac{\Gamma(-\frac{1}{2}+A)}{\Gamma(\frac{1}{2}-A)}\left(-\frac{A}{4\omega}\right)^{\frac{1}{2}}.
\end{equation}
As a result, this limit is valid only if $A<0$, with the growth rate scaling as  
\begin{equation}
\label{scale2}
\hat{\gamma} \sim S^{-1}.
\end{equation}

\subsection{$\omega \approx |A| \ll 1$ limit, $A>0$}
In this limit, $\omega$ is comparable to $A>0$. Thus, we can Taylor expand the Gamma functions given in equation (\ref{sim_dr}) around $A/\omega=1$  to obtain  
\begin{equation}
    \label{limit3}
\left({\frac{\hat{\gamma}}{S\hat{k}^{2}}}\right)^{\frac{1}{4}}\frac{1}{A+\omega}\frac{\Gamma(-\frac{1}{2}+A)}{\Gamma(\frac{1}{2}-A)}\frac{\omega}{A-\omega}.
\end{equation}
In order to have  a finite value in the limit $S\rightarrow\infty$, one must therefore impose that
\begin{equation}
    A^2\sim \omega^{2},
\end{equation}
such that the growth rate scaling becomes  
\begin{equation}
\label{scale3}
\hat{\gamma} \sim S^{-1/3}.
\end{equation}

This regime does not exist for $A<0$, for which equation (\ref{sim_dr}) tends to zero in the limit of $S\rightarrow\infty$ and thus a solution to equation (\ref{dr}) for the growth rate does not exist.   

\acknowledgements{We acknowledge support by DoE grant DE-FG02-04ER54742 and NSF award 2108320. We wish to thank Scott J. Hopper for useful discussions on the solution of the layer equations.}

\bibliographystyle{jpp}
\bibliography{jpp-instructions}

\end{document}